\documentclass[twocolumn,10pt]{IEEEtran}
\usepackage{ifpdf, flushend,subfigure}

\ifCLASSINFOpdf
  \usepackage[pdftex]{graphicx}
  \graphicspath{{../pdf/}{../jpeg/}}
  \DeclareGraphicsExtensions{.pdf,.jpeg,.png}
\else
  \usepackage[dvips]{graphicx}
  \graphicspath{{../eps/}}
  \DeclareGraphicsExtensions{.eps}
\fi
\usepackage[cmex10]{amsmath}
\usepackage {amssymb}
\usepackage{algorithmic}
\usepackage{array}
\usepackage{mdwmath}
\usepackage{mdwtab}
\usepackage{eqparbox}
\usepackage{url}
\usepackage{hyperref}
\usepackage{algorithm}
\usepackage{algorithmic}

\newcommand{\argmin}{\operatornamewithlimits{argmin}}

\newcommand{\beq}{\begin{equation}}
\newcommand{\eeq}{\end{equation}}
\newcommand{\beqn}{\begin{eqnarray}}
\newcommand{\eeqn}{\end{eqnarray}}
\newcommand{\beqno}{\begin{eqnarray*}}
\newcommand{\eeqno}{\end{eqnarray*}}
\newcommand{\bma}{\begin{displaymath}}
\newcommand{\ema}{\end{displaymath}}
\newcommand{\bnu}{\begin{enumerate}}
\newcommand{\enu}{\end{enumerate}}
\newcommand{\bce}{\begin{center}}
\newcommand{\ece}{\end{center}}
\newcommand{\btb}{\begin{tabular}}
\newcommand{\etb}{\end{tabular}}

\begin{document}

%
\title{Compressed Sensing Based Data Processing and MAC Protocol Design for Smartgrids}

\author{\IEEEauthorblockN{Le Thanh Tan and Long Bao Le}  
\thanks{The authors are with INRS-EMT, University of Quebec,  Montr\'{e}al, Qu\'{e}bec, Canada. 
Emails: \{lethanh,long.le\}@emt.inrs.ca.}}


\maketitle

\begin{abstract}
\boldmath
In this paper, we consider the joint design of data compression and 802.15.4-based medium access control (MAC) protocol for smartgrids with renewable energy.
We study the setting where a number of nodes, each of which comprises electricity load and/or renewable sources, report periodically their injected powers to 
a data concentrator. Our design exploits the correlation of the reported data in both time and space
to perform efficient data compression using the compressed sensing (CS) technique and efficiently engineer
the MAC protocol so that the reported data can be recovered reliably within minimum reporting time. Specifically,
we perform the following design tasks: i) we employ the two-dimensional (2D) CS technique to compress
the reported data in the distributed manner; ii) we propose to adapt the 802.15.4 MAC protocol frame structure
to enable efficient data transmission and reliable data reconstruction; and iii) we develop an analytical model
based on which we can obtain the optimal parameter configuration to minimize the reporting delay.
Finally, numerical results are presented to demonstrate the effectiveness of our design.
\end{abstract}

\IEEEpeerreviewmaketitle

\section{Introduction}
\label{Intro}

Future energy grid is expected to integrate more distributed and renewable
energy resources with significantly enhanced communications infrastructure 
for timely and reliable data exchanges between the control center and various grid control points \cite{Stojmenovic14}.
The advanced communications infrastructure supports many critical grid control, monitoring, and management
operations and emerging smartgrid applications. The communications infrastructure is typically
hierarchical, e.g., data communications between customers and local concentrators via field/neighborhood area networks, and 
between local concentrators and the utility company via long-haul wide area networks \cite{ERDF, Sendin12}. 
The former is usually based on the low bandwidth communications technologies such as Zigbee, WiFi, and power line communications (PLC) while 
the later is required to have higher capacity, which can be realized by employing LTE, 3G cellular, WiMAX, and fiber optics for example .

Our work concerns the design of data compression and MAC protocol for the field/neighborhood area network where PLC is employed to report 
injected powers from grid connection points to the local concentrator. 
In fact, several smartgrid projects in France \cite{ERDF}, and Spain \cite{Sendin12} have chosen PLC for smartgrid deployment since
PLC can be realized with low-cost modems and it utilizes available electricity wires for data communications.
We focus on the reporting of injected powers at different grid connection points since this information can be employed for many grid applications
such as line-failure prediction \cite{Sendin13} or congestion management and various grid control applications \cite{Lo12}.
Furthermore, the utility control center can utilize collected data to further estimate the complete phasor data at different nodes which can be then used
in the control of voltages and reactive powers or in the active load management \cite{Alam13, Samarakoon11},  
and outage management \cite{Liu02}.

There have been some existing works that study data compression and MAC protocol design issues in the smartgrid and wireless network contexts.
In addition, there are two popular standards for the PLC technology, namely PRIME and G3-PLC \cite{Matan13a}--\cite{Sendin11}.
In \cite{Alam13}, the authors study the state estimation problem where the voltage phasors at different nodes are recovered based on limited
reported data. The authors in \cite{Fazel11} consider random access exploiting the CS capability for energy efficiency communications in wireless sensor networks.
However, joint design of communications access and data compression for smartgrid is still very under-explored by the existing literature.

In this paper, we propose to engineer the PRIME MAC protocol jointly with CS-based data compression for smartgrid communications networks.
Specifically, we consider a distributed random reporting (DRR) mechanism where each grid connection point (referred to as a node in the sequel) reports
 its injected power data in a probabilistic manner and using the PRIME MAC protocol for data transmission.
The Kronecker CS technique (2D CS), which can exploit the spatio-temporal correlation of data, is then employed for data reconstruction at 
the utility control center. We develop an analytical model for the proposed design based on which we can guarantee reliable data construction. In addition,
 we present an algorithm which determines efficient configuration for MAC parameters so that the average reporting time is minimized. 

 
\section{System Model}
\label{SystemModel}

\begin{figure}[!t]
\centering
\includegraphics[width=90mm]{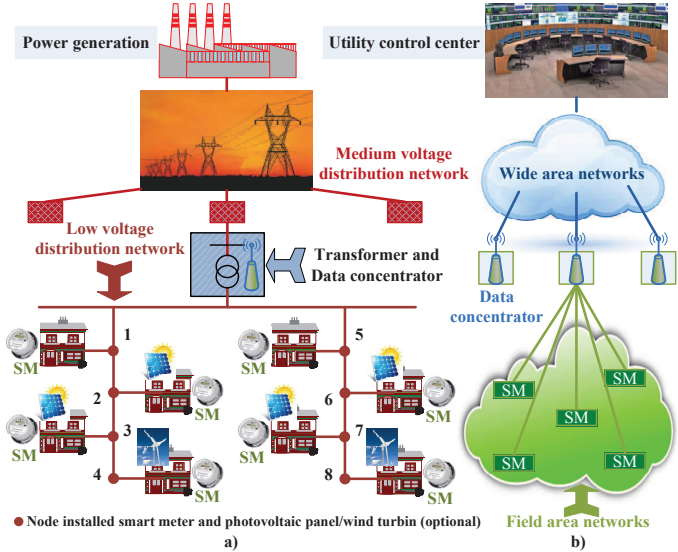}
\caption{Smartgrids with advanced communications infrastructure}
\label{LV_network}
\end{figure}

We consider the one-hop communications between a group of $n_S$ grid nodes and one data concentrator (point to multi-points).
We assume that each node is equipped with a smart meter (SM) that reports 
injected power data to the data concentrator using PLC.\footnote{We implicitly
assume that the wide area network has very high bandwidth and can deliver the
received data at the data concentrators to the utility control center without errors}. 
Moreover, each node is assumed to comprise a load and/or a solar/wind generator.
Large-scale deployment of such renewable generators at customer  premises, the light poles, or advertising panels
 has been realized for urban areas in Europe \cite{ERDF, Sendin12}. In reality,
the data concentrator is installed at the same location as the transformer to collect data from grid nodes of the distribution network.
This communication and grid model is shown in Fig.~\ref{LV_network}.

We assume that each node must report the injected power value once in each reporting interval (RI) \cite{Alam13, Samarakoon11}.
For most available smart meters, the RI can be configured from a few minutes to hours \cite{SMII}.
Let $\mathbf{S}_{l,i}$, $\mathbf{S}_{g,i}$ and $\mathbf{S}_i$ be the load, distributed generation and injected powers at node $i$, respectively.
Then, the injected power at node $i$ can be expressed as $\mathbf{S}_i = \mathbf{S}_{g,i} - \mathbf{S}_{l,i}$.
We assume that the control center wishes to obtain information about $\mathbf{S}_i$ for all nodes $i (i \in \left[1, n_S\right])$
in each RI.\footnote{The proposed framework can be applied to other types of grid data as long as they exhibit sparsity in the
time and/or space domains.}
Under the constraint of low bandwidth properties of PLC 
\cite{Matan13a}--\cite{Sendin11}, our objectives are to develop joint design of the data compression 
and MAC protocol so that reliable data reporting can be achieved within minimum reporting time (RT). 
One important design requirement target is that
the MAC protocol is distributed which allows low-cost and large-scale deployment.

It has been shown in some recent works that power data related to power grids with renewable energy exhibit
strong correlation (implying sparsity) over space and time \cite{Glasbey08}, \cite{Soares08}. 
Therefore, to realize the underlying compression benefits,  the control center only needs to collect the reported data in $m_T < n_T$ RIs (i.e., compression over time) 
from a subset of $n_S$ nodes with size $m_S < n_S $ (i.e., compression over space) to reconstruct the complete data
for $n_S$ nodes and $n_T$ RIs. Note that the reconstructor can be at the data concentrator or at the control center. 
The later is chosen in this work as the best choice for reducing 
bandwidth usage in the second phase.
We can easily extend to apply our proposed method to the applications where each node would report their load and 
distributed generation powers instead of injected powers. 
Because these load and distributed generation powers also have strong correlation as presented above.

For practical implementation, the data transmission and construction can be performed in a rolling
manner where data construction performed at a particular RI utilizes the reported data over the latest $n_T$ RIs.
To guarantee the desirable data construction quality, the control center must receive sufficient data which
is impacted by the employed reporting mechanism and MAC protocol. Specifically, we must determine the values of 
$m_T$ and $m_S$ for some given values $n_T$ and $n_S$ to achieve the desirable data construction reliability.

\section{CS-Based Data Compression}
\label{GCSMAC}

\subsection{CS-Based Data Processing}

Without loss of generality, we consider data construction for one data field for $n_T$ RIs and $n_S$ nodes.
Let $\mathbf{Z}$ be an $n_S \times n_T $ matrix whose $(i,j)$-th element denotes the injected
power at node $i$ and RI $j$. We will refer to the data from one RI (i.e., one column of $\mathbf{Z}$)
as one data block in this paper. 
From the CS theory, we can compress the data if they possess 
sparsity properties in a certain domain such as wavelet domain. Specifically, we can express the
data matrix $\mathbf{Z}$ as
\beqn \label{wavtran}
\mathbf{Z} = \mathbf{\Psi}_S \mathbf{A} \mathbf{\Psi}_T^T
\eeqn
where  $\mathbf{\Psi}_S \in \mathbb{R}^{n_S \times n_S} $ and $\mathbf{\Psi}_T \in \mathbb{R}^{n_T \times n_T}$
denote wavelet bases in space and time dimensions, respectively \cite{Duar12}.  The sparsity of $\mathbf{Z}$ can be
observed in the wavelet domain if matrix $\mathbf{A}$ has certain $K$ significant (nonzero) coefficients where $K <n_S \times  n_T$.

We now proceed to describe the data compression and reconstruction operations. Let us denote
 $\mathbf{\Phi}_S \in \mathbb{R}^{m_S \times n_S} $ (for space) and $\mathbf{\Phi}_T \in \mathbb{R}^{m_T \times n_T}$ (for time) as the two 
sparse observation matrices where entries in these two matrices are i.i.d uniform random numbers where $m_S <  n_S $ and $m_T < n_T $.
Specifically, we can employ  $\mathbf{\Phi}_S$  and $\mathbf{\Phi}_T$
to sample the power data from which we obtain the following observation matrix
\beqn
\mathbf{Y} = \mathbf{\Phi}_S \mathbf{Z} \mathbf{\Phi}_T^T.
\eeqn

Let $N_{\Sigma} = n_T n_S $ and $M = m_T m_S$ be the number of elements of $\textbf{Z}$ and $\mathbf{Y}$, respectively.
From the CS theory, we can reliably reconstruct the data matrix $\textbf{Z}$ by using the observation matrix $\mathbf{Y}$ if 
$m_T$ and $m_S$ are appropriately chosen. For the smartgrid communications design, this implies that
the control center only needs to collect $M$ injected power elements instead of $N_{\Sigma}$ values
for reliable construction of the complete data field. 

We now describe the data construction for $\textbf{Z}$ by using the observation matrix $\mathbf{Y}$.
To achieve this, the control center can determine matrix $\mathbf{A}$, which corresponds
to wavelet transform of the original data $\textbf{Z}$ as described in (\ref{wavtran}),
by solving the following optimization problem
\beqn
\min_ \textbf{A} \left\|\textbf{A}\right\|_2 \,\,\,
\text{s.t.}  \,\,\, \left\|\textbf{vec}\left(\mathbf{Y}\right) - \mathbf{\bar Y}\right\|_2 \leq \epsilon \label{OPT_RE_EQN}
\eeqn
where $\mathbf{\bar Y} = \left(\mathbf{\Phi}_S \otimes \mathbf{\Phi}_T \right) \left(\mathbf{\Psi}_S \otimes \mathbf{\Psi}_T \right) \textbf{vec}\left(\textbf{A}\right)$, $\otimes$ is the Kronecker product, $\textbf{vec} \left(\mathbf{X}\right)$ denotes the vectorization of the matrix $\mathbf{X}$ formed by stacking the rows of $\mathbf{X}$ into a single column vector. We can indeed solve problem (\ref{OPT_RE_EQN}) by using the Kronecker CS algorithm \cite{Duar12} to obtain $\mathbf{A}^* = \argmin_\textbf{A}  \left\|\textbf{A}\right\|_2$.
Then, we can obtain the estimation for the underlying data as $\textbf{vec}\left(\textbf{Z}^* \right) = \mathbf{\Psi}_S \otimes \mathbf{\Psi}_T \mathbf{A}^*$.
More detailed discussions of this data reconstruction algorithm can be found in \cite{Duar12}. 

Now there are two questions one must answer to complete the design: 1) how can one choose $m_S $ and $m_T$
to guarantee reliable data reconstruction for the underlying data field?; and 2) how can one design the data sampling
and MAC protocol so that the control center can have sufficient information for data reconstruction? We will provide
the answers for these questions in the remaining of this paper.

\subsection{Determination of $m_S$ and $m_T$}
\label{GCSMAC11}

\begin{figure}[!t]
\centering
\includegraphics[width=60mm]{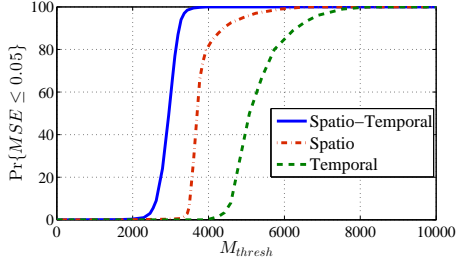}
\caption{Probability of success vs $M_{\sf thresh} $ with $n_S = 128 $ and $n_T = 128 $.}
\label{Novel_CDF_Numnode128_Numsample128}
\end{figure}

\begin{table}
\centering
\caption{Calculation of $M_{\sf thresh} $, $m_S $ and $m_T $}
\label{table}
\scriptsize
\setlength{\tabcolsep}{5pt}
\begin{tabular}{|c|c|c|c|c|c|c|}
\hline 
($n_S $, $n_T $) & (64,64) & (64,128) & (64,256) & (128,128) & (128,256) & (256,256)\tabularnewline
\hline
\hline 
$M_{\sf thresh} $ & 1551 & 1892 & 2880 & 3196 & 4620 & 9200 \tabularnewline
\hline 
($m_S $, $m_T $) & (33,47) & (22,86) & (16,180) & (47,68) & (30,154) & (80,115)\tabularnewline
\hline
\end{tabular}
\end{table} 

We would like to choose $m_S$ and $m_T$ so that $M = m_S m_T$
is minimum. Determination of the optimal values of $m_S$ and $m_T$ turns out to be a non-trivial task \cite{Duar12}.
So we propose a practical approach to determine $m_S$ and $m_T$. It is intuitive
that $m_S$ and $m_T$ should be chosen according to the compressibility in the space and time dimensions, respectively.
In addition, these parameters must be chosen so that the reliability of the data reconstruction meets the predetermined requirement,
which is quantified by the mean square error (MSE).

To quantify compressibility in the space and time, we consider two other design options, viz. temporal CS and spatial CS alone. 
For the former, the control center reconstructs data for any particular node by using the observations of only that node (i.e., we ignore spatial correlation).
For the later, the control center reconstructs the data in each RI for all nodes without exploiting the correlation over different RIs.
For fair quantification, we determine the MSE for one data field (i.e., for data matrix $\mathbf{Z}$ with $n_S \times n_T $ elements) for these two design options
where $MSE = \left\|\mathbf{Z}-\mathbf{Z}^*\right\|_2^2/\left\|\mathbf{Z}\right\|_2^2$.

For each spatial CS and temporal CS cases, we generate 1000 realizations of the injected powers based on which we perform 
the data reconstruction using the 1D CS for different
values of $m_S$ and $m_T$, respectively. Then, we obtain the empirical probability of success for the data reconstruction
versus $M_{S} = m_S n_T$ and $M_{T} = n_S m_T$, respectively
where the ``success'' means that the MSE is less than the target MSE.
From the obtained empirical probability of success, we can find the required values of $M_S$ and $M_T$, which are denoted as $M_{S,{\sf thresh}}$ and $M_{T,{\sf thresh}}$, respectively, 
to achieve the target success probability.
Having obtained $M_{S,{\sf thresh}}$ and $M_{T,{\sf thresh}}$ capturing the compressibility in the
space and time as described above, we choose $m_S $ and $m_T$ for the 2D CS so that 
$m_S/m_T = n_S/n_T \times M_{S,{\sf thresh}}/M_{T,{\sf thresh}}$.
Similarly, we obtain the empirical probability of success for the 2D CS based on which 
we can determine the minimum value of $M_{\sf thresh} = m_S m_T$ to achieve the target success probability.

To obtain numerical results in this paper, we choose target $MSE = 0.05$ and target success probability equal 0.95.
Fig.~\ref{Novel_CDF_Numnode128_Numsample128} shows the empirical probability of success versus $M_{{\sf thresh}}= m_S m_T$ for 
$n_S = n_T = 128$ where  $M_{{\sf thresh}}$
in the horizontal axis represents the $M_{S,{\sf thresh}}$, $M_{T,{\sf thresh}}$, and $M_{{\sf thresh}}$ for 1D and 2D CS for simplicity.
The data model for the injected power will be described
in Section~\ref{Data_Model}. For this particular setting, we can obtain the ratio  $m_S/m_T = 0.691 $
from which can obtain the values of $m_S = 47 $ and $m_T = 68 $.

Similarly, we determine the $M_{\sf thresh}$, $m_S $ and $m_T$ for different scenarios with the corresponding $(n_S, n_T)$ in Table~\ref{table}.
For all cases, we can observe that $m_S < n_S$ and $m_T < n_T$, which demonstrates the benefits of performing data compression using the 2D CS. 
Having determined $m_S $ and $m_T$ as described above, the remaining tasks are to design
the DDR mechanism and MAC protocol that are presented in the following.

\section{DDR And MAC Protocol Design}
\label{Net_des_Per_Ana}

\subsection{Distributed Data Reporting Design}

In any RI, to perform reconstruction for the data field corresponding to the latest $n_T$ RIs, 
the control center must have data in $m_T$ RIs, each of which comprises $m_S$ injected powers
from $m_S$ nodes. With the previously-mentioned rolling implementation, the control center
can broadcast a message to all nodes to request one more data block (the last column of data matrix $\textbf{Z}$) if it has only $m_T-1$ data blocks 
to perform reconstruction in the current RI. At the beginning, the control center
can simply send $m_T$ broadcasts for $m_T$ randomly chosen RIs out of $n_T$ RIs and it performs data construction
at RI $n_T$ upon receiving the requested data blocks.

\subsection{MAC Protocol Design}

\begin{figure}[!t] 
\centering
\includegraphics[width=85mm]{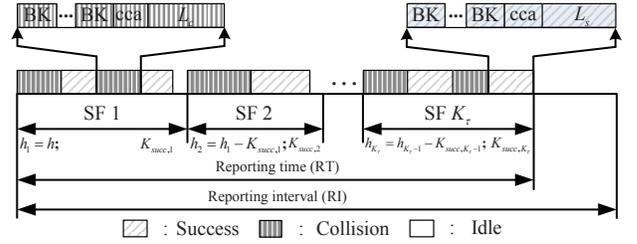}
\caption{Timing diagram of the MAC superframe structure for one RI.}
\label{cycletime}
\end{figure}

If there is a broadcast message from the control center, we assume that each node participates in the contention process with probability $p_s$
using the slotted CSMA/CA MAC protocol in any RI. The slotted CSMA/CA MAC protocol \cite{Prime} with the proposed frame structure
is employed for data transmissions as described in the following. We set the optionally contention-free period to zero since we focus on distributed access design in this work.
Moreover, we assume there are $K_{\tau} $ superframes (SFs) in the contention period of any RI where $SF_i  = SF_0 \times 2^{BO_i}$ is the length of SF $i $,
 $SF_0 $ is the base length of SF, $BO_i \in \left[0,BO_{\sf max}\right]$ is the beacon order. Therefore, the reporting time (RT) in the underlying RI
is $ \sum_{i=1}^{K_{\tau}} SF_i $. We will optimize the parameters  $K_{\tau} $ and $\left\{BO_i\right\}$ to minimize the RT
while guaranteeing the desirable data reconstruction quality later. The superframe structure in one RI is illustrated in Fig.~\ref{cycletime}.

In each SF, the nodes that choose to access the channel (with probability $p_s$) perform the contention using the standardized slotted CSMA/CA protocol as follows.
A contending node randomly chooses a number in $\left[0, W_0\right]$ as the backoff counter ($W_0 \!=\! 2^{priority}$) and starts counting down.
If the counter reaches zero, the node will perform the clear channel  assessment (CCA) for $priority$ times (we set $priority \!=\! 2$). 
It will transmit data and wait for ACK if all CCAs are successful.
The reception of ACK is interpreted as a successful transmission, otherwise this is a collision.
In the case of failure in any CCA, the node attempts to perform backoff again with doubling backoff window.
In addition, each node is allowed to access channel up to $NB\!+\!1$ times. Since the length of the SF
is limited, a node may not have enough time to transmit its data and ACK packets at the end of the last SF.
In this case, we assume that the node will wait for the next SF to access the channel.
We refer to this as the deference state in the following.

\subsection{MAC Parameter Configuration for Delay Minimization}
\label{Net_design} 

We consider optimizing the MAC parameters $K_{\tau}, p_s, \left\{BO_i\right\}$ to minimize the reporting time in each RI
by solving the following problem:
\beqn
\label{OPT_DOST_EQN}
\begin{array}{l}
 {\mathop {\min }\limits_{K_{\tau}, p_s, BO_i}} \quad \mathcal{D}(K_{\tau}, p_s, \left\{BO_i\right\}) = \sum_{i=1}^{K_{\tau}} SF_0 \times 2^{BO_i} \\
 \mbox{s.t.} \quad \Pr\left\{K_{\text{succ}} \geq m_{S} \right\} \geq P_{\text{suff}},  \\
 \quad \quad  0 \leq K_{\tau} \leq K_{\tau,{\sf max}}, 0 \leq BO_i \leq BO_{\sf max}, 0 \leq p_s  \leq 1. \\
 \end{array}
\eeqn

In (\ref{OPT_DOST_EQN}), the first constraint means that the control center must receive $m_{S}$ packets (i.e., $m_{S}$ injected power values from $m_{S}$ nodes) 
with probability $P_{\text{suff}} \approx 1$. Note that one would not be able to deterministically guarantee the reception of $m_{S}$ packets due to the
random access nature of the DDR and MAC protocol.
The probability in this constraint can be written as
\beqn 
\Pr\left\{K_{\text{succ}} \geq m_{S} \right\} \!\! = \!\! \sum_{h= m_{S} }^{n_S} \!\! \Pr\left\{\widehat{m}_S = h\right\} \!\!\!\!\!\! \sum_{K_{\sf succ} = m_{S} }^h \sum_{l =1 }^{\left|\Xi\right|}  \prod_{i=1}^{K_{\tau}} \label{Prob_mthresh_OST_EQN_1}\\
\times \sum_{K_{S,i} = K_{{\sf succ},i}}^{K_{S,i,{\sf max}}} \!\!\!\!\!\!\!\! \Pr\!\left\{K_i = K_{S,i} \left|h_i \right.\right\} \! \Pr\!\left\{{\bar K}_i = K_{\text{succ},i} \left|K_{S,i},h_i \right.\right\} \label{Prob_mthresh_OST_EQN_3}
\eeqn
where $K_{\text{succ}}$ denotes the number of successfully transmitted packets in the RI and $\Pr\!\left\{\widehat{m}_S \!=\! h\right\} $ is the probability of $h$ nodes joining the contention.
Since each node decides to join contention with probability $p_s$, $\Pr\!\left\{\widehat{m}_S \!=\! h\right\} $ is expressed as 
\beqn
\label{Prob_htrans}
\Pr\left\{\widehat{m}_S = h\right\} = \left(\begin{array}{*{20}{c}} {n_S} \\  {h}  \\ \end{array}\right) p_s^h \left(1-p_s\right)^{n_S - h}.
\eeqn
In (\ref{Prob_mthresh_OST_EQN_1}) and (\ref{Prob_mthresh_OST_EQN_3}), we consider all possible scenarios so that the total number of
successfully transmitted packets over $K_{\tau}$ SFs is equal to  $K_{\text{succ}}$ where $K_{\text{succ}} \in \left[m_{S}, h\right]$. Here,
$K_{\text{succ},i}$ denotes the number of successfully transmitted packets in SF $i$ so that we have $\sum_{i=1}^{K_{\tau}} K_{\text{succ},i} = K_{\text{succ}}$.
In particular, we generate all possible combinations of $\left\{K_{\text{succ},i}\right\} $ for $K_{\tau} $ SFs and $\Xi $ represents the set of all possible combinations ($\left|\Xi\right|$ is the number of possible combinations). 

For each combination, we calculate the probability that the control center receives  $K_{\text{succ}}$ successful packets.
Note that  a generic frame may experience
one of the following events:  success, collision, CCA failure and deference.
Also, there are at most $K_{S,i,{\sf max}}$ frames in any SF $i$ where $K_{S,i,{\sf max}} = \left\lfloor SF_i/\min\left\{ NB+1, L_s+2 \right\}\right\rfloor $ since
the smallest length of a CCA failure frame is $NB +1 $ slots while the minimum length of a successful frame is $L_s+2$ where
$L_s$ is the required time for one successful transmission and 2 represents the two CCA slots.
We only consider the case that $K_{{\sf succ},i} \leq K_{S,i} \leq K_{S,i,{\sf max}}, \forall i \in \left[1, K_{\tau}\right]$.

In (\ref{Prob_mthresh_OST_EQN_3}), $\Pr\! \left\{K_i \!=\! K_{S,i} \left|h_i \right.\right\}$ is the probability that there are $K_{S,i}$ generic frames in SF $i$ given that $h_i$ nodes join
contention where $h_1 \!=\!h$ and $h_i \!=\! h_{i-1}\!-\! K_{\text{succ},i-1}$ since successfully transmitting node will not perform contention in the following frames.
Moreover, $\Pr\!\left\{{\bar K}_i \!=\! K_{\text{succ},i} \left|K_{S,i},h_i \right.\right\}$ is the probability that $K_{\text{succ},i}$ nodes transmit successfully in SF $i$ 
given that there are $h_i$ contending nodes and $K_{S,i}$ generic frames.

In order to calculate $\Pr\left\{K_i = K_{S,i} \left|h_i \right.\right\}$ and $\Pr\left\{{\bar K}_i = K_{\text{succ},i} \left|K_{S,i},h_i \right.\right\}$,
we have to analyze the Markov chain capturing detailed operations of the MAC protocol. 
For simplicity, we set $NB_i = NB = 5$, which is the default value. The analysis of the Markov chain model is omitted due to the space constraint. 
Then we determine $\Pr\left\{K_i = K_{S,i} \left|h_i \right.\right\} $ and $\Pr\left\{{\bar K}_i = K_{\text{succ},i} \left|K_{S,i},h_i \right.\right\} $ as follows.

\subsubsection{Calculation of $\Pr\left\{K_i = K_{S,i} \left|h_i \right.\right\} $}
\label{Pr_K_S_i1} 

In SF $i $ there are $h_i $ contending nodes and $K_{S,i}$ generic frames where frame $j$ has length $T_{ij}$.
We can approximate the distribution of generic frame length $T_{ij}$ as the normal distribution.
So the probability of having $K_{S,i} $ generic frames is written as
\beqn
\label{K_genfram_EQN}
\Pr\!\left\{\!K_i \!= \!K_{S,i} \left|h_i\right. \!\right\} \!\!=\!\! \Pr\{\sum_{j=1}^{K_{S,i}} \! T_{ij} \!\! = \!\! SF_i \} \!=\!\mathcal{Q} (\frac{SF_i\!-\!K_{S,i} {\bar T}_i}{\sqrt{K_{S,i} \sigma^2_i}})
\eeqn
where ${\bar T}_i $ and $\sigma^2_i $ are the average and variance of the generic frame length, respectively whose
calculations are presented in Appendix~\ref{average_variance}.

\subsubsection{Calculation of $\Pr\left\{{\bar K}_i = K_{\text{succ},i} \left|K_{S,i},h_i \right.\right\} $}
\label{Pr_K_S_i} 
The second quantity, $\mathcal{P} = \Pr\left\{{\bar K}_i = K_{\text{succ},i} \left|K_{S,i},h_i \right.\right\} $ is equal to
\beqn
\mathcal{P}  = \!\!\!\!\!\! \sum_{j = 0}^{K_{S,i}- K_{\text{succ},i}} \sum_{k= 0}^{1} \left(\!\!\!\begin{array}{*{20}{c}} {K_{S,i}} \\ { K_{\text{succ},i},j,k,l}  \\ \end{array} \!\!\!\right) \mathcal{P}_{\text{succ},h_i}^{ K_{\text{succ},i}}  \mathcal{P}_{\text{coll},h_i}^j \mathcal{P}_d^k \mathcal{P}_{\text{ccas},h_i}^{l} \label{P_2_OST_EQN}
\eeqn
where $K_{\text{succ},i}\!+\!j\!+\!k\!+\!l \!=\! K_{S,i}$; $j$, $k$, and $l$ represent the number of frames with collision, deference, and CCA failure, respectively. Moreover,
$\mathcal{P}_{\text{succ},h_i}$,  $\mathcal{P}_{\text{coll},h_i}$, $\mathcal{P}_{\text{ccas},h_i}$, and $\mathcal{P}_d$
denote the probabilities of success, collision, CCA failure, and deference, respectively, whose calculations are given in Appendix~\ref{average_variance}. 
In (\ref{P_2_OST_EQN}), we generate all possible combinations each of which has different numbers of success, collision, CCA failure, and deference frames. Also,
the product behind the double summation is the  probability of one specific combination.

\subsection{MAC Parameter Configuration Algorithm}

\begin{algorithm}[h]
\caption{\textsc{Optimization of MAC Parameters}}
\label{OPT_Delay}
\begin{algorithmic}[1]

\FOR {each value of $K_{\tau} \in [1,K_{\tau,{\sf max}}]$}

\FOR {each possible set $\left\{BO_i\right\}$}

\STATE Find optimal ${\bar p}_s $ as ${\bar p}_s = \mathop {\argmin} \limits_{0 \leq p_s \leq 1} \mathcal{D} \left( K_{\tau}, \left\{BO_i\right\}, p_s\right) $.

\ENDFOR

\STATE The best $\left(\left\{\bar{BO}_i\right\}, {\bar p}_s\right)$ for each $K_{\tau} $ is $\left(\left\{\bar{BO}_i\right\}, {\bar p}_s\right) = \mathop {\argmin} \limits_{\left\{{ BO}_i\right\}, {\bar p}_s} \mathcal{D} \left(K_{\tau}, \left\{BO_i\right\}, {\bar p}_s\right) $.
\ENDFOR

\STATE The final solution $\left( {\bar K}_{\tau}, \left\{\bar{BO}_i\right\}, {\bar p}_s  \right) $ is determined as $\left( {\bar K}_{\tau}, \left\{\bar{BO}_i\right\}, {\bar p}_s  \right) = \mathop {\argmin} \limits_{K_{\tau}, \left\{\bar{BO}_i\right\}, {\bar p}_s} \mathcal{D} \left(K_{\tau}, \left\{\bar{BO}_i\right\}, {\bar p}_s\right) $.

\end{algorithmic}
\end{algorithm}

Since there are only finite number of possible choices for $K_{\tau} \in [1,K_{\tau,{\sf max}}]$ and the set $\left\{{BO}_i\right\}$,
we can search for the optimal value of $p_s$ for given $K_{\tau}$ and $\left\{{BO}_i\right\}$ as in step 3.
Then, we search over all possible choices of $K_{\tau}$ and the set $\left\{{BO}_i\right\}$ to determine the optimal
configuration of the MAC parameters (in steps 5 and 7).

\vspace{10pt}
\section{Performance Evaluation and Discussion}
\label{Results}

\subsection{Data Modeling and Simulation Setting}
\label{Data_Model}

Since real data for load and renewable are not available, we synthetically generate them by using the existing methods \cite{Glasbey08}, \cite{Soares08}.
We set the RI equal 5 minutes and employ the auto-regressive models \cite{Soares08} to synthesize the data. 
Specifically, we utilize the autoregressive process as a time series model
which consists of the deterministic and stochastic components \cite{Soares08} as
$ X_t = X_t^d + X_t^s $ where $t$ is the time index,  $X_t^d$ and $X_t^s$ denote the deterministic and stochastic components of $X_t$, respectively. 
Here, $X_t$ $X_t^d$, and $X_t^s$ are commonly used to represent active load power $P_l$, and reactive load power $Q_l$. 
The deterministic component which depicts the trend of data is represented by the trigonometric functions \cite{Soares08} as follows: 
\beqn
\label{deter_EQN}
X^d_t = \chi_0 + \sum_{i=1}^{m_h} \left(\chi_{\text{re},i} \text{sin}\left(\frac{2\pi k_i t}{288}\right)+ \chi_{\text{im},i}\text{cos}\left(\frac{2\pi k_i t}{288}\right)\right)
\eeqn
where $m_h$ is the number of harmonics (i.e., the number of trigonometric functions) 
where $\chi_{\text{re},i}$ and $\chi_{\text{im},i}$ are the coefficients of the harmonics, 
$\chi_0$ is the constant, $k_i \leq 288/2$ for $\forall i \in [1, m_h]$.
The stochastic component is modeled by the first order autoregressive process AR(1) and expressed as \cite{Soares08}
$X_{t+1}^s = \varphi_t X_t^s + U_t$
where $\varphi_t$ is the AR(1) coefficient and $U_t$ is the white noise process with zero mean and variance of $(1-\varphi_t)$.

The data set of the active and reactive load powers from the years 2006 to 2010 is used to estimate
the parameters ($\chi_0$, $\chi_{\text{re},i}$, $\chi_{\text{im},i}$, $m_h$, $\varphi_{t}$) \cite{UCIMLR}. 
We first use the ordinary least squares (OLS) to estimate $\chi_0$, $\chi_{\text{re},i}$, and $\chi_{\text{im},i}$, 
$i \in [1, m_h]$ where Bayesian information criterion (BIC) \cite{Soares08} is used to determine $m_h$ 
significant harmonics. Then we also use the OLS algorithm to estimate $\varphi_t$ \cite{Soares08}. 
Note that these parameters are revised and stored in every 5-minute interval over one day. 
The same time series model described above is used to obtain distributed generation power $S_g$.
Then the spatial correlation of distributed generation powers is captured as in \cite{Glasbey08} 
where the correlation coefficient between 2 distributed generators $i$ and $j$ is $\rho_{i,j} = \exp\left(-d_{i,j}/d\right)$. 
Here $d = 20 km$ and $d_{i,j}$ is the distance between generators $i$ and $j$, which is randomly chosen in $\left(0,1 km\right]$.

In the following simulation, we assume that wind/solar generators are installed at half of considered nodes for all following experiments.
We ignore the node index $i$ in these notations for brevity.
The RI is set as $\tau_T = 5$ minutes. The target probability in the constraint (\ref{OPT_DOST_EQN}) is chosen as $P_{\text{suff}} = 0.9$.
The MAC parameters are chosen as  $L_s = T_p + t_{ACK} + L_{ACK}$, $T_p = 5+L_{MAC}$ slots ($L_{MAC} = 2$ is the MAC header), $L_{ACK} = 2$ slots, $t_{ACK} = 1$ slot, $t_{ACK,ti} = 4$ slots
where $T_p $ is the length of packet, $t_{ACK} $ is the idle time before the ACK, $L_{ACK} $ is the length of ACK, $t_{ACK,ti}$ is the timeout of the ACK. For all the results
presented in this section, we choose $n_T=256$.

\subsection{Numerical Results and Discussion}

\begin{figure}[!t]
\centering
\mbox{\subfigure[]{\includegraphics[width=1.7in]{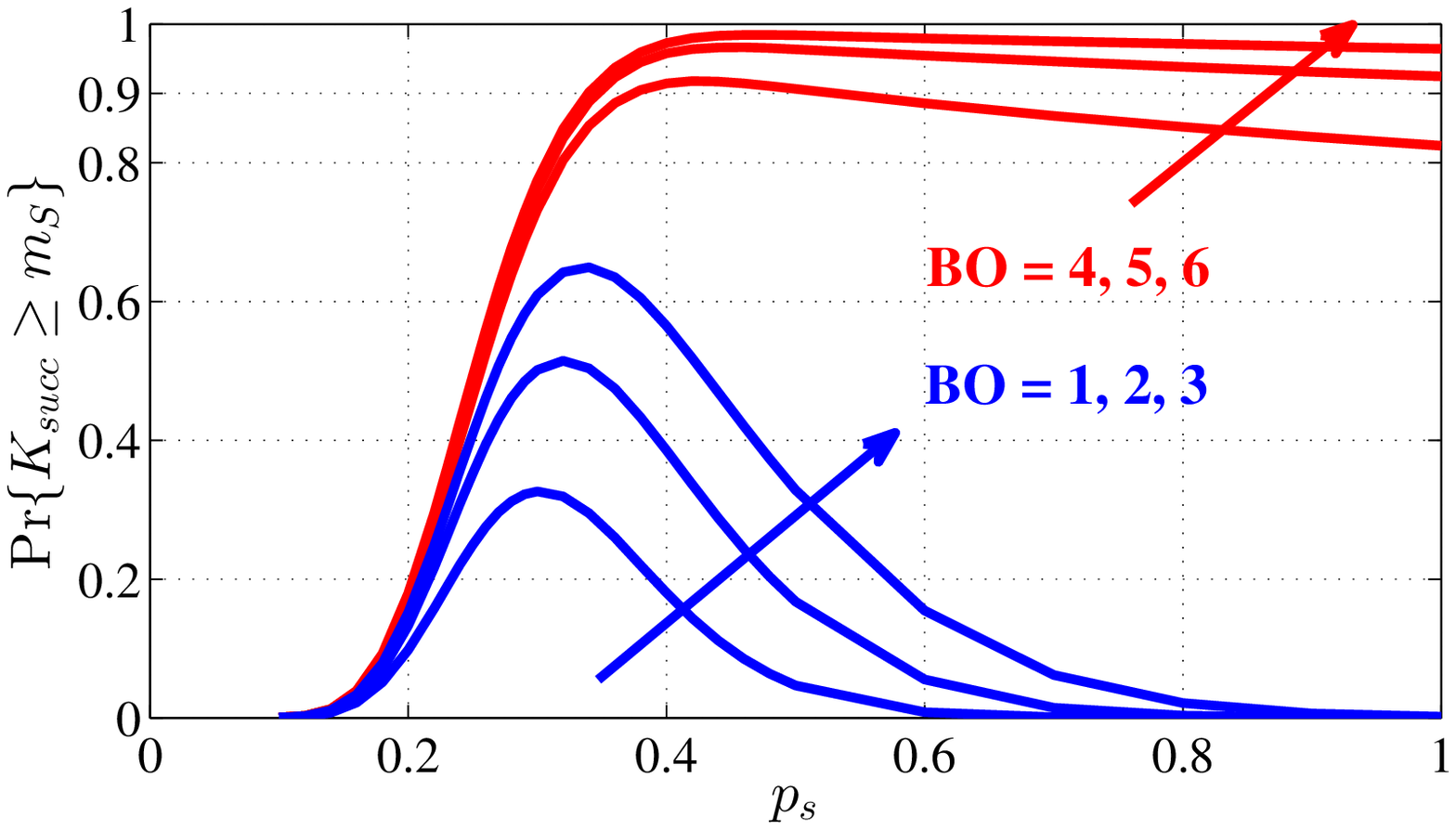} \label{Prsuff_Numnode_64_K_tau_3_BO_p_s1‎}}  
\subfigure[]{\includegraphics[width=1.7in]{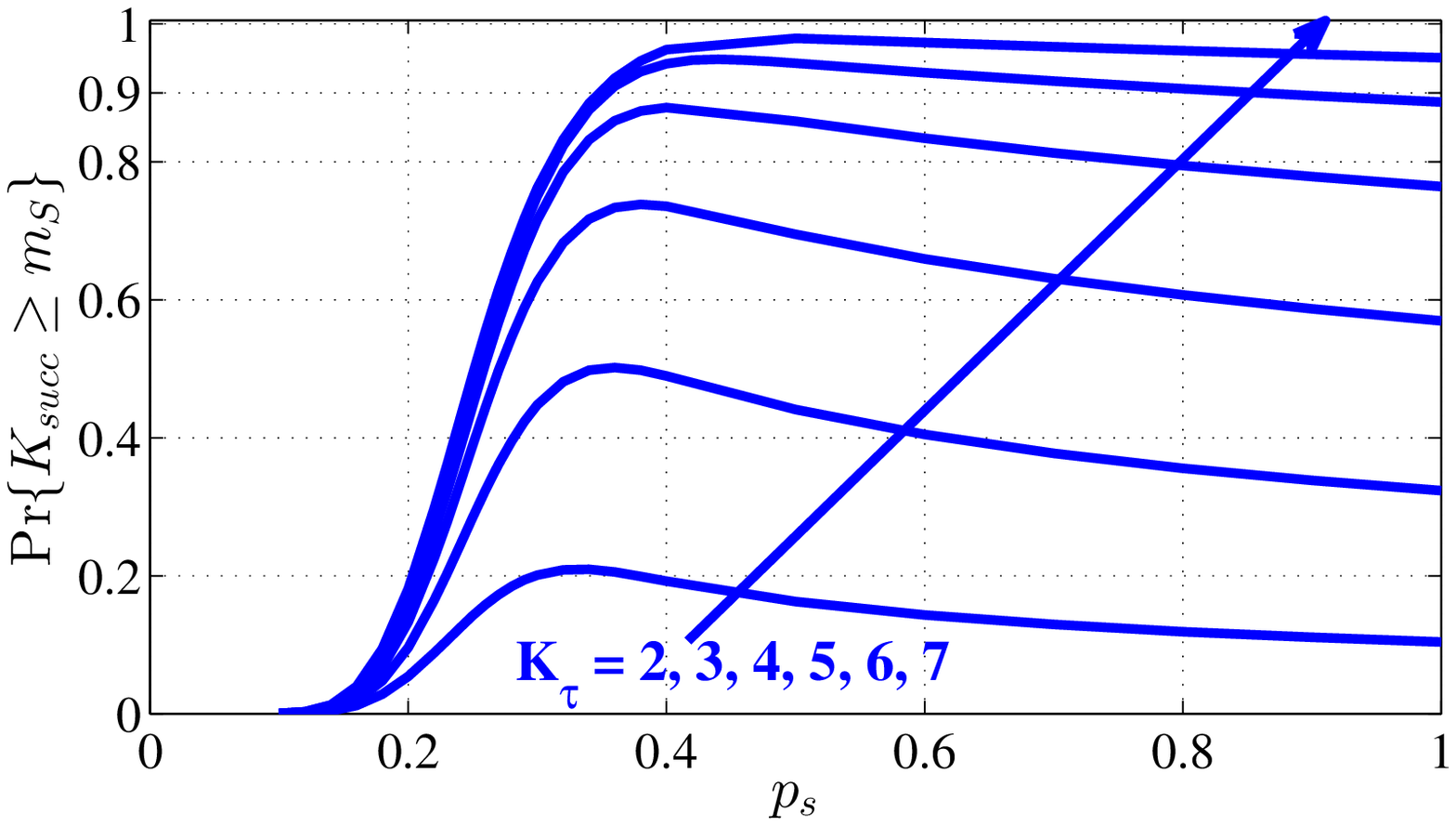} \label{Prsuff_Numnode_64_K_tau_BO_3_p_s2}} }
\caption{$\Pr\!\left\{K_{\text{succ}} \! \geq \! m_{S}\right\} $ vs. $p_s$ for (a) $(n_S, m_{S}, K_{\tau}) \!=\! (64, 16, 3)$,
 various values of $BO$,  (b) $(n_S, m_{S}, BO) = (64, 16, 3)$, various values of $K_{\tau}$.}
\end{figure}




\subsubsection{Sufficient Probability}

In Fig.~\ref{Prsuff_Numnode_64_K_tau_3_BO_p_s1‎}, we show the variations of $\Pr\!\left\{K_{\text{succ}} \!\geq\! m_{S}\right\}$ versus 
$p_s$ for different values of $BO_i \!=\! BO $ (i.e., all SFs employ the same $BO$) where $m_{S} \!=\! 16$, $K_{\tau} \!=\! 3$,
 and $n_S \!=\! 64$. It can be observed that there exists an optimal $p_s$ that maximizes $\Pr\!\left\{K_{\text{succ}} \!\geq\! m_{S}\right\}$ for any value of $BO$.
This optimal value is in the range $\left[0.3, 0.5\right]$. Furthermore, 
the $BO$ must be sufficiently large ($BO \!\geq\! 4$) to meet the required target value $P_{\text{suff}}$=0.9. In addition, larger values of $BO$ lead
 to longer SF length, which implies that more data packets can be transmitted.
In Fig.~\ref{Prsuff_Numnode_64_K_tau_BO_3_p_s2}, we show the probability $\Pr\left\{K_{\text{succ}} \geq m_{S}\right\} $ versus $p_s $ for
different values of $K_{\tau} $ where we set $n_S = 64 $ and $BO = 3$. This figure confirms that the maximum $\Pr\left\{K_{\text{succ}} \geq m_{S}\right\} $
becomes larger with increasing $K_{\tau} $ where we can meet the target probability $P_{\text{suff}}$=0.9 as $K_{\tau} \geq 6$.


\begin{figure}[!t]
\centering
\mbox{\subfigure[]{\includegraphics[width=1.7in]{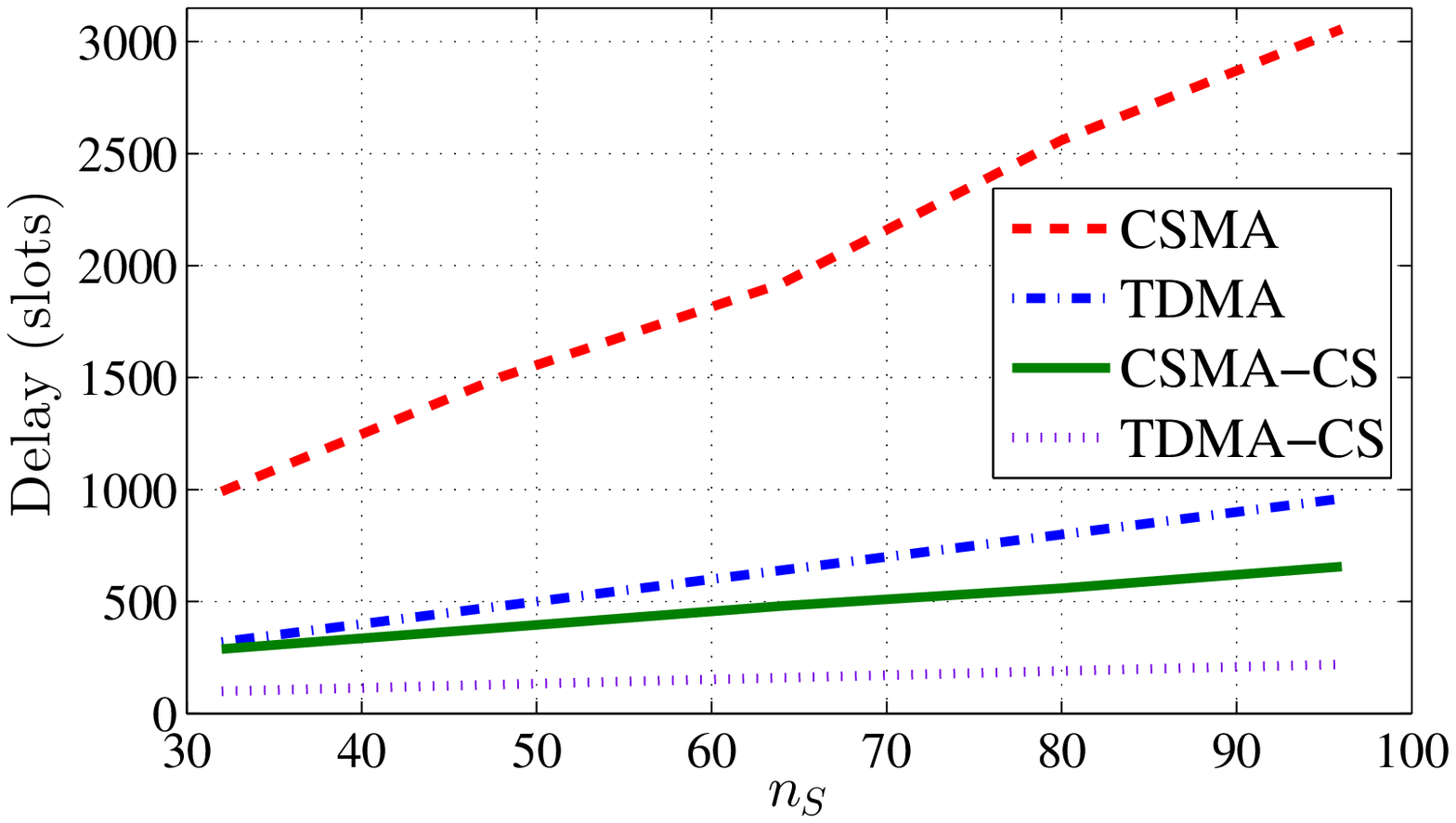} \label{Delay_ns}}  
\subfigure[]{\includegraphics[width=1.7in]{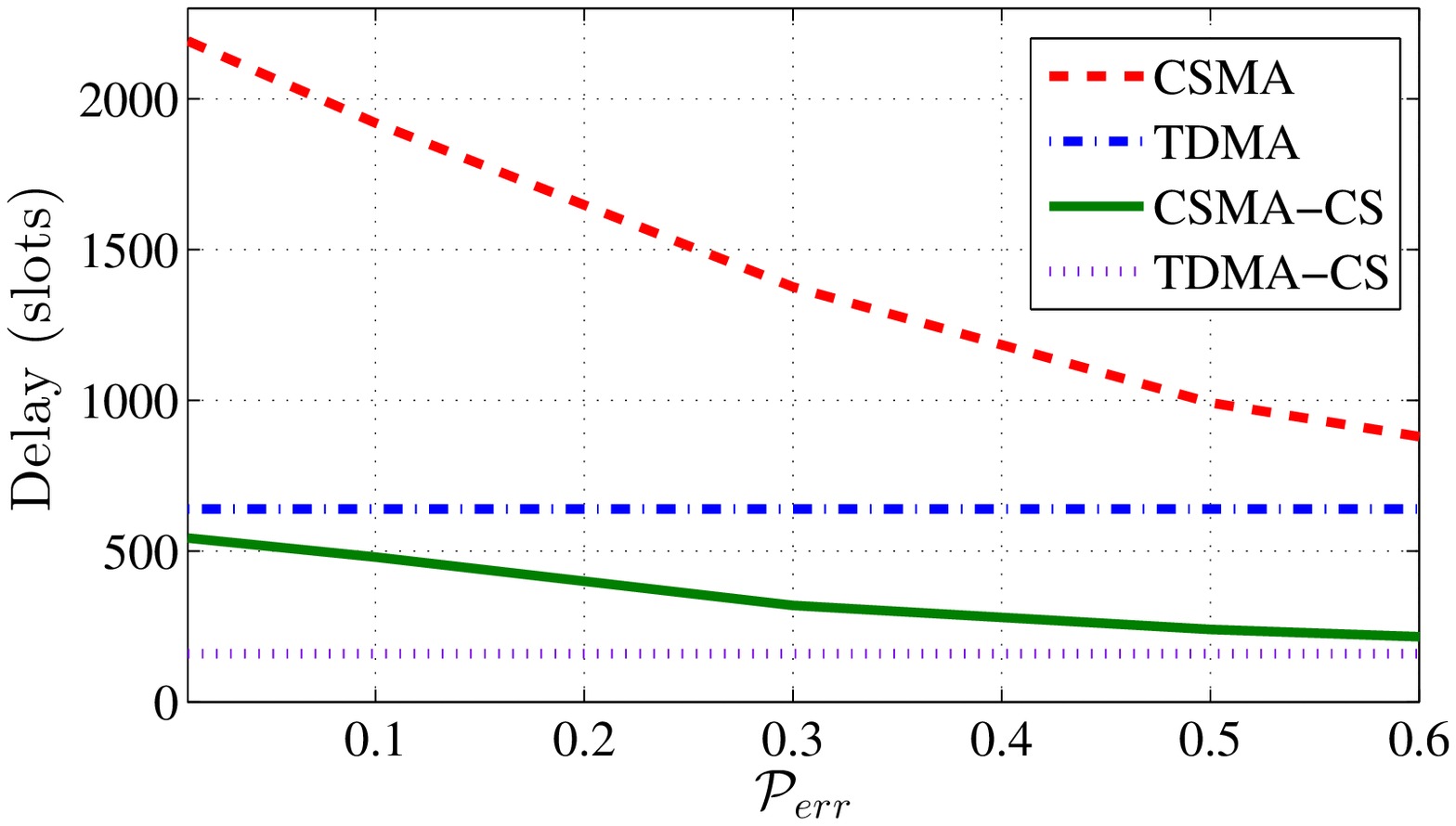} \label{Delay_P_err_n_S_64}}}
\caption{Reporting delay $\mathcal{D}$ vs. (a) $n_S$,  (b) $\mathcal{P}_{err}$.}
\end{figure}

%
%
%

\subsubsection{Reporting Delay}

We now show the optimal reporting delay $\mathcal{D}$ in one RI versus the number of nodes $n_S$ for different schemes,
namely TDMA, CSMA, TDMA-CS, and our proposed CSMA-CS schemes in Fig.~\ref{Delay_ns}. Here, X-CS refers 
to scheme X that integrates the CS capability and X refers to scheme X without CS.
Moreover, TDMA is the centralized non-contention time-division-multiple-access MAC, which always achieves better performance CSMA MAC.
For both CSMA and CSMA-CS schemes, their MAC parameters are optimized by using Alg.~\ref{OPT_Delay}.
It can be seen that our proposed CSMA-CS protocol achieves much smaller delay than the CSMA scheme, which
confirms the great benefits of employing the CS. In addition, TDMA-CS outperforms
our CSMA-CS protocol since TDMA is a centralized MAC while CSMA is a randomized distributed MAC.
Finally, this figure shows that our CSMA-CS protocol achieves better delay performance than 
the TDMA scheme. 

We illustrate the variations of the optimal reporting delay with $\mathcal{P}_{err} $ for different schemes where 
$\mathcal{P}_{err} \!=\! 1\!-\!\Pr\!\left\{K_{\text{succ}} \!\geq\! m_{\text{S}}\right\}$
and $n_S \!=\! 64$ in Fig.~\ref{Delay_P_err_n_S_64}. This figure shows that as $\mathcal{P}_{err} $ increases, the reporting delay decreases.
This indeed presents the tradeoff between the reporting delay and data reconstruction error $\mathcal{P}_{err}$. 
Note that the delay of TDMA-CS protocol is the lower bounds for all other schemes. Interestingly, as $\mathcal{P}_{err}$ increases
the delay gap between the proposed CSMA-CS and the TDMA-CS schemes become smaller. 


\begin{figure}[!t]
\centering
\mbox{\subfigure[]{\includegraphics[width=1.7in]{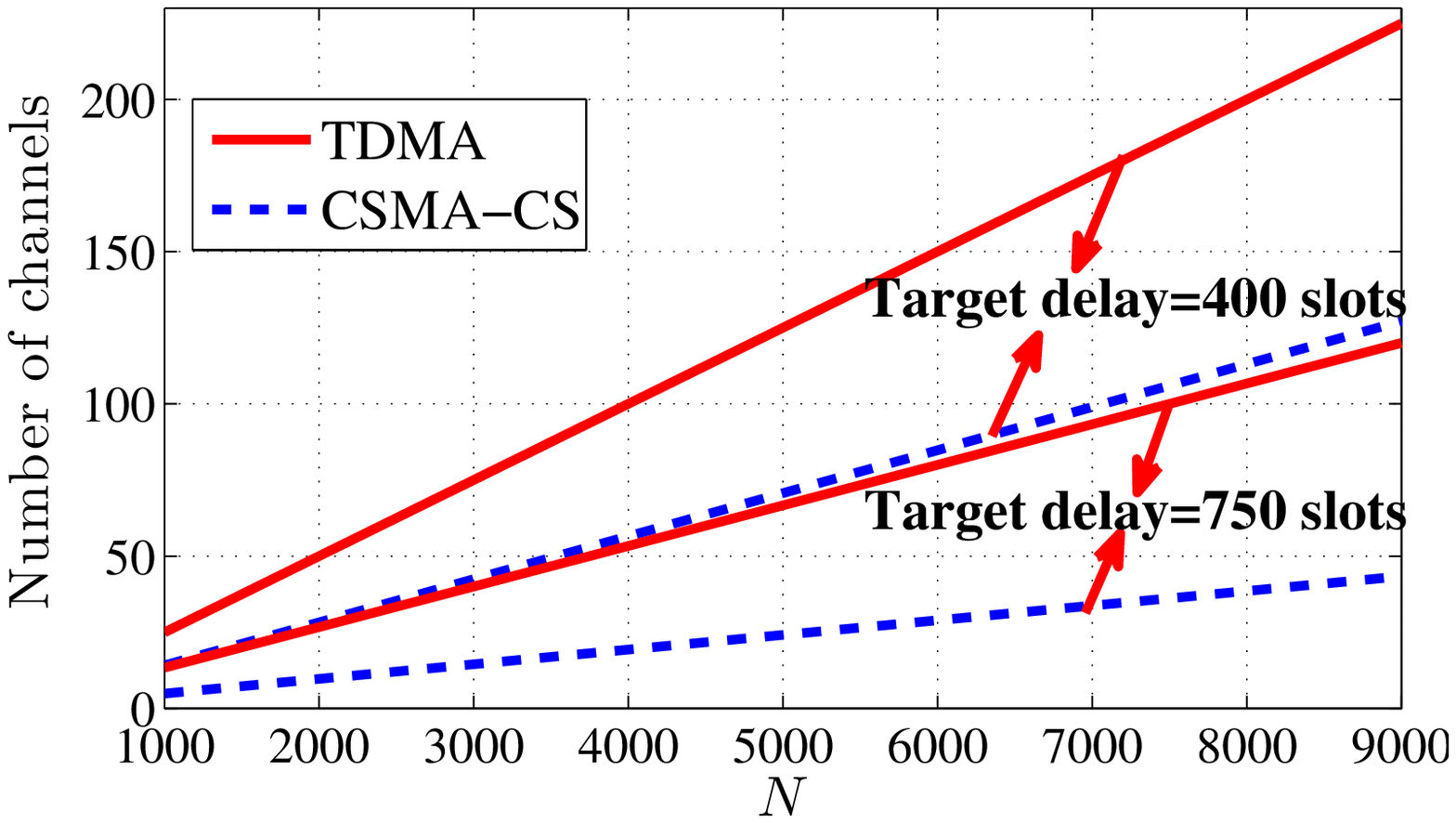} \label{Numchan_N_Numnode_BOp_s}}  
\subfigure[]{\includegraphics[width=1.7in]{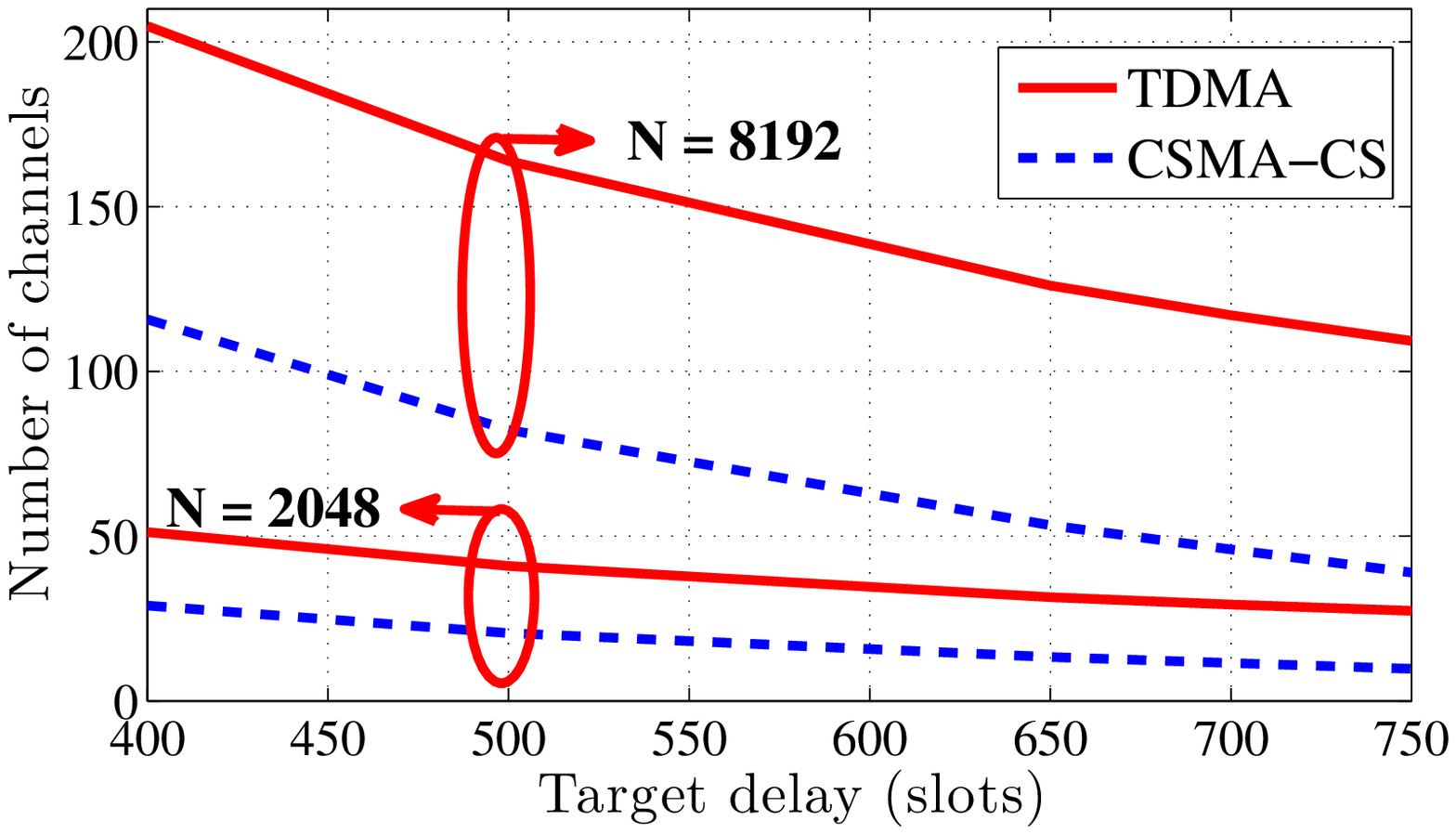} \label{Delay_target_Numnode}}}
\caption{Number of channels vs. (a) $N$,  (b) target delay.}
\end{figure}

%
%

\subsubsection{Bandwidth Usage}

To ease the exposition, we do not show the 
performance of bandwidth usage for the CSMA and TDMA-CS schemes. 
We consider a particular neighborhood with $N>n_S$ nodes, whose simultaneous transmissions
can collide with one another. Suppose these $N$ nodes want to report data to a control center. In this case, we would need
$N/n_S$ orthogonal channels\footnote{We ignore the fact that this number must be integer for simplicity} to support these communications. Suppose that 
the considered smartgrid application has a maximum target delay of $\mathcal{D}_{\sf max}$ determined by using the results in Fig.~\ref{Delay_ns}.
We should design the group size with $n_S$ nodes as large as possible while respecting this target delay.

Let the maximum numbers of nodes for one group under TDMA and CSMA-CS schemes while still respecting the target delay be $n_S^{\sf TDMA}$ and $n_S^{\sf CSMA-CS}$, respectively.
Note that the TDMA scheme uses all $n_T$ RIs for data transmission while
 the CSMA-CS scheme only chooses $m_T$ RIs for each $n_T$ RIs to transmit the data.
Thus, the CSMA-CS scheme allows $n_T/m_T$ groups to share one channel for each interval of $n_T$ RIs.
As a result, the number of channels needed for $N$ nodes is $N/n_S^{\sf TDMA}$ for the TDMA scheme
and $N/n_S^{\sf CSMA-CS} \times m_T/n_T$ for the CSMA-CS scheme.
Specifically,  we have $n_S^{\sf TDMA}\!\! = \left\{40, 75\right\}$ and $n_S^{\sf CSMA-CS}\!\! = \left\{55, 128\right\}$ for TDMA and CSMA-CS schemes
for the target delay values of $\mathcal{D}_{\sf max} = \left\{400, 750\right\}$ slots, respectively.
Then we calculate the required bandwidth (i.e., number of channels) for TDMA and CSMA-CS schemes with a given number of nodes $N$.

In Fig.~\ref{Numchan_N_Numnode_BOp_s}, we show the required number of channels versus $N$.
It can be observed that our proposed CSMA-CS scheme requires less than half of the bandwidth demanded by the TDMA scheme.
Also when the network requires smaller target delay, we need more channels for both schemes as expected.
Finally, Fig.~\ref{Delay_target_Numnode} describes the variations in the number of channels
 versus the target delay for $N = \left\{2048, 8192\right\}$.
Again, our proposed CSMA-CS scheme provides excellent bandwidth saving compared to the TDMA scheme.

\section{Conclusion}
\label{Conclusion} 

We have proposed the joint design of data compression using the CS techniques and CSMA MAC protocol for smartgrids
with renewable energy. Then, we have presented the design and optimization of the MAC protocol to minimize the reporting
delay. Numerical results have confirmed the significant performance gains of the proposed design compared to other non-compressed solutions.

\appendices

\section{Calculation of ${\bar T}_i $ and $\sigma^2_i $}
\label{average_variance} 

In this appendix, we determine $\bar{T}_i $ and $\sigma^2_i $ for a particular SF $i$.
For simplicity, we again omit the index $i$ and $h_i$ in all related parameters when this does not create confusion.
First, we can express the probability generating function (PGF) of the generic frame, $T(z) $ which includes success, collision, CCA failure and deference,
as
\beqn
T\left(z\right) = \mathcal{P}_{\sf succ} T_S\left(z\right) + \mathcal{P}_{\text{coll}} T_C\left(z\right) + \mathcal{P}_{\text{ccas}} T_F\left(z\right) + \mathcal{P}_d T_D\left(z\right)
\eeqn
Here we denote $\mathcal{P}_{\text{ccas}} $, $\mathcal{P}_{\text{coll}} $ and $\mathcal{P}_{\text{succ}} $ as the probabilities of CCA failure, collision and success, respectively. These probabilities can
be calculated as $\mathcal{P}_{\text{ccas}} = \left(1-\mathcal{P}_d\right) \lambda^{NB+1} $, $\mathcal{P}_{\text{coll}} = p_c \left(1-\mathcal{P}_d\right) \left(1-\lambda^{NB+1}\right) $, and $\mathcal{P}_{\sf succ} = 1-\mathcal{P}_{\text{coll}}- \mathcal{P}_{\text{ccas}}- \mathcal{P}_d $, where $p_c = 1-\left(1-\phi\right)^{h-1} $.
Moreover, we also denote $T_S\left(z\right) $, $T_C\left(z\right) $, $T_F\left(z\right)$ and $T_D\left(z\right) $ as the PGFs of durations of success, collision, CCA failure and deference, respectively. 
These quantities can be calculated as in \cite{Park10}.

Finally, we can determine $\bar T $ and $\sigma^2$ from the first and second derivation of $T\left(z\right) $ at $z = 1 $, i.e.,
\beqn
\bar T = \frac{dT}{dz}\left(1\right) ; 
\sigma^2 = \frac{d^2T}{dz^2}\left(1\right) + \bar T - \left(\bar T\right)^2. \label{var_genfram_EQN}
\eeqn
These parameters $\bar T $ and $\sigma^2$ will be utilized in (\ref{K_genfram_EQN}).

\bibliographystyle{IEEEtran}


\end{document}